\documentclass[12pt,a4paper]{article}
\usepackage{epsf}
\newcommand{\be}{\begin{equation}}
\newcommand{\ee}{\end{equation}}
\newcommand{\bea}{\begin{eqnarray}}
\newcommand{\eea}{\end{eqnarray}}
\newcommand{\sgn}{\,{\rm sgn}}
\begin{document}
\title{Tiles and colors}
\author{Bernard Nienhuis \\
Instituut voor Theoretische Fysica\\
Valckenierstraat 65\\
1018 XE Amsterdam\\
Netherlands\\
e-mail  nienhuis@wins.uva.nl}
\maketitle
\begin{abstract}
Tiling models are classical statistical models in which different 
geometric shapes, the tiles, are packed together such that they cover 
space completely. 
In this paper we discuss a class of two-dimensional tiling 
models in which the tiles are rectangles and isosceles triangles. Some 
of these models have been solved recently by means of Bethe Ansatz. 
We discuss the question why only these models in a larger family  
are solvable, and we search for the Yang-Baxter structure 
behind their integrablity.
In this quest we find the Bethe Ansatz solution of the problem of
coloring the edges of the square lattice in four colors, such that 
edges of the same color never meet in the same vertex.
\end{abstract}
\newpage
This paper is dedicated to Professor Rodney Baxter for his 
great contributions to statistical mechanics, and is particularly
inspired by his lectures and seminars, 
where he so admirably shares with 
the audience his own fascinations and interests.

\section{Lattices and tilings}
In many models for solid state physics the lattice as a
simple idealization of the crystal is a given ingredient.
Of course in reality the crystal is the result of the interactions and 
statistics of the particles in the system.
The ideal lattice can be viewed as a periodic repetition of the unit cell.
This makes it periodic and it is typically also 
symmetric under a specific discrete rotational symmetry group.
The compatibility of the translational and rotational symmetry
gives the well known crystallographic restrictions on the possible 
rotational symmetry groups.

The quasicrystal is a state of matter which does not observe this 
restriction. 
This is possible only because it is not strictly periodic.
Instead it has a property called quasiperiodicity.
A quasiperiodic function in $d$ dimensions is a restriction of a
periodic function in a dimension higher than $d$ 
to a $d$-dimensional hyperplane.
The analogue for quasicrystals of the ideal lattice is the
quasilattice or perfect quasiperiodic tiling.
It is built as a repetition of more than one `unit cell' also called tile.
Examples of these structures are the famous Penrose 
tilings\cite{penrose} with rhombi as tiles, or  their three-dimensional 
analogues with Ammann rhombohedra\cite{ammann,rt1}.
In contrast to lattices, these quasilattices may have non-crystallographic 
rotational symmetries, reflecting the genuinely crystallographic 
symmetries of the periodic functions in higher dimensions.
However, these non-crystallographic symmetries are not true 
symmetries of the function itself, but only of the absolute value of its 
Fourier transform. 
Because it is precisely this quantity that is measured 
in diffraction experiments, the symmetry is nevetheless quite real.

\section{Random tilings as statistical models}
It was proposed by Elser\cite{elser} that these noncrystallographic 
symmetries can be achieved not only by a quasiperiodic arrangement, 
but also by an ensemble of arbitrary packings of the tiles.
This happens in the same way that the high temperature phase of a 
statistical model reflects the symmetry of the hamiltonian, not in any 
particular configuration, but only in the complete ensemble and in 
statistical averages.
For instance, in the case of the Penrose 
rhombi, the angles between the edges are all multiples of $\pi/5$. 
Therefore if the plane is covered by copies of these rhombi, then also all 
the edges in the whole tiling have angles with one another which are 
multiples of $\pi/5$. As a consequence the continuous rotational
symmetry of the plane, is reduced to at most a ten-fold discrete symmetry.
The complete ensemble of tilings with these tiles 
will naturally have this tenfold rotational symmetry, 
unless the symmetry is further reduced by 
possible symmetry breaking schemes.
This observation has led to the 
study of what is now known as random tilings\cite{rt1,rt2}. 
Despite their name these models have no intrinsic randomness.
These are discrete 
statistical models of which the configurations are tilings of space by 
means of a limited set of tiles. They are called random only to emphasize
the difference with perfect quasiperiodic tilings.

In principle one may introduce an 
interaction energy between adjacent tiles, but in most cases studied 
there is no other interaction than the full packing constraint, i.e.
that the tiles cover space without holes or overlaps.

Though the phrase of random tilings came up in the study of
quasicrystals, many random tilings actually live on the lattice
\cite{rtl}.
The dimer prob\-lem\cite{dimers} is a typical example, but also e.g. the
hard hexagon model\cite{hh}
can be viewed as a tiling with hexagons and triangles.
This paper, however, is concerned with random tilings of which the
tiles do not fit together on a regular lattice.
\section{Solvable random tilings}
\begin{figure}
\setlength{\unitlength}{7mm}
\centerline{
\begin{picture}(20,7)(0,0)
\put(0 ,0){\makebox(6,6){\epsfxsize=45mm \epsfbox{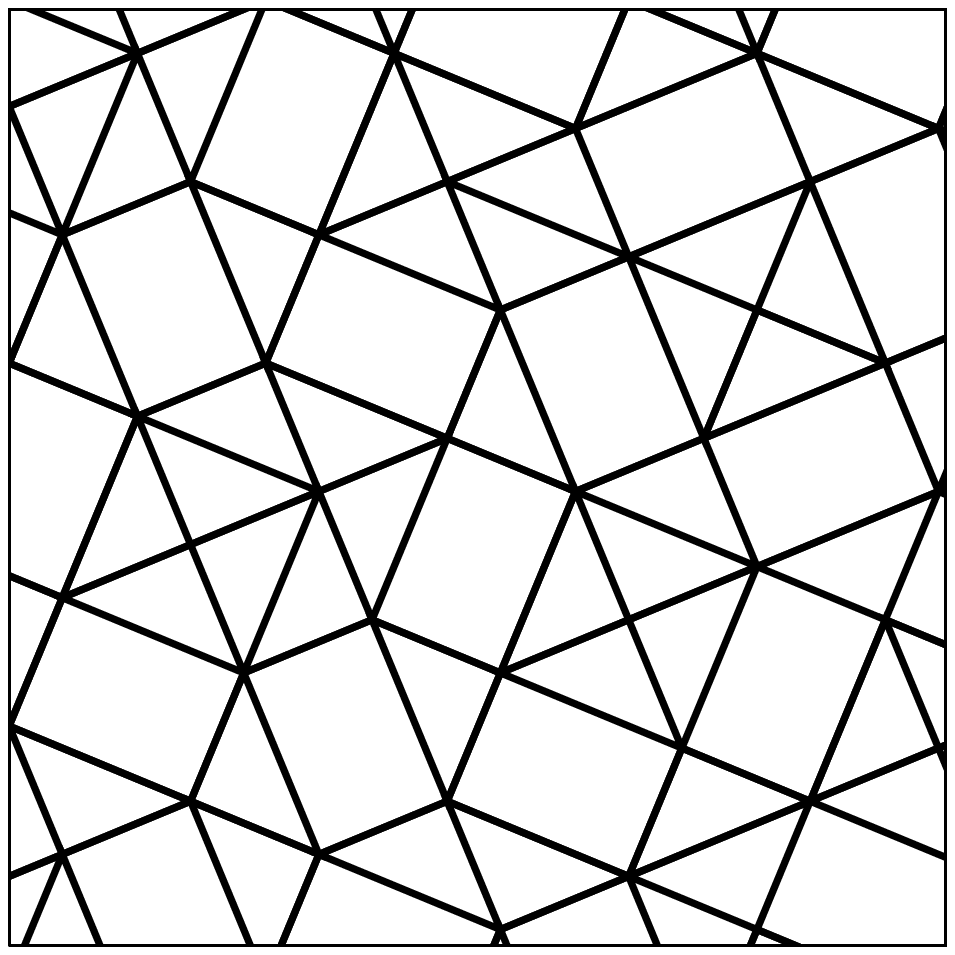} } }
\put(7 ,0){\makebox(6,6){\epsfxsize=45mm \epsfbox{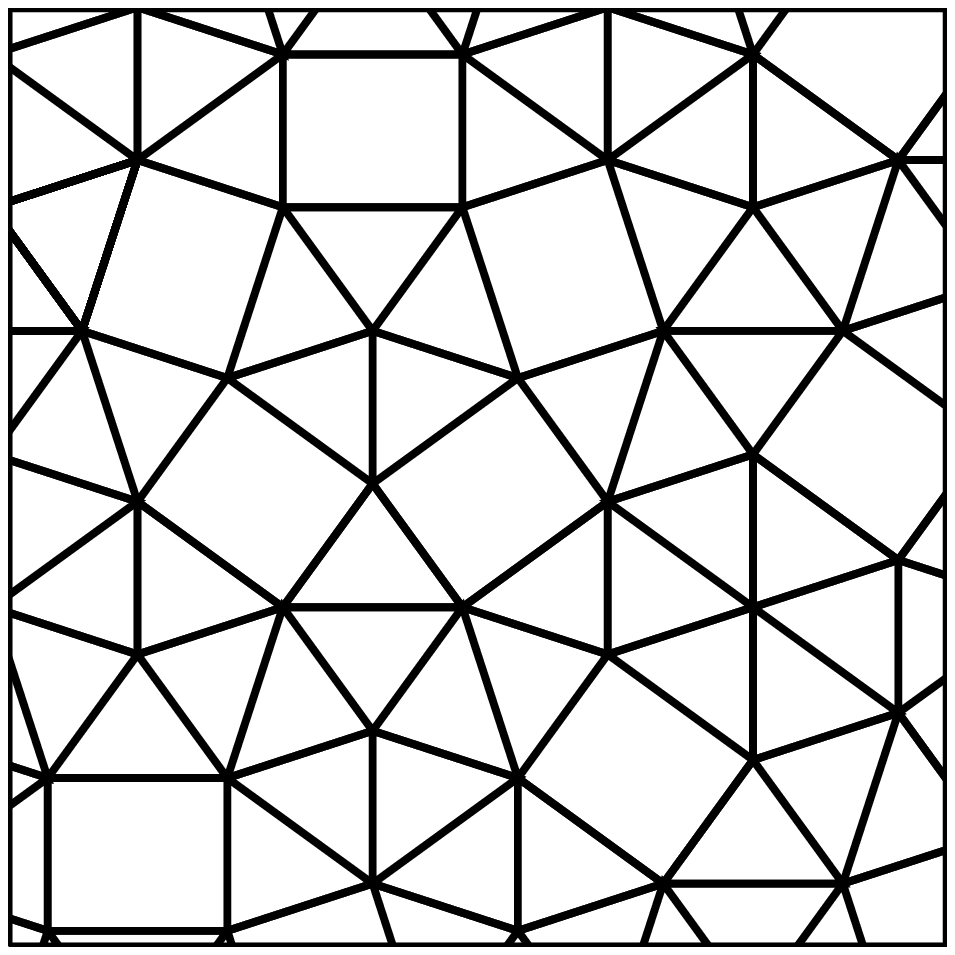} } }
\put(14,0){\makebox(6,6){\epsfxsize=45mm \epsfbox{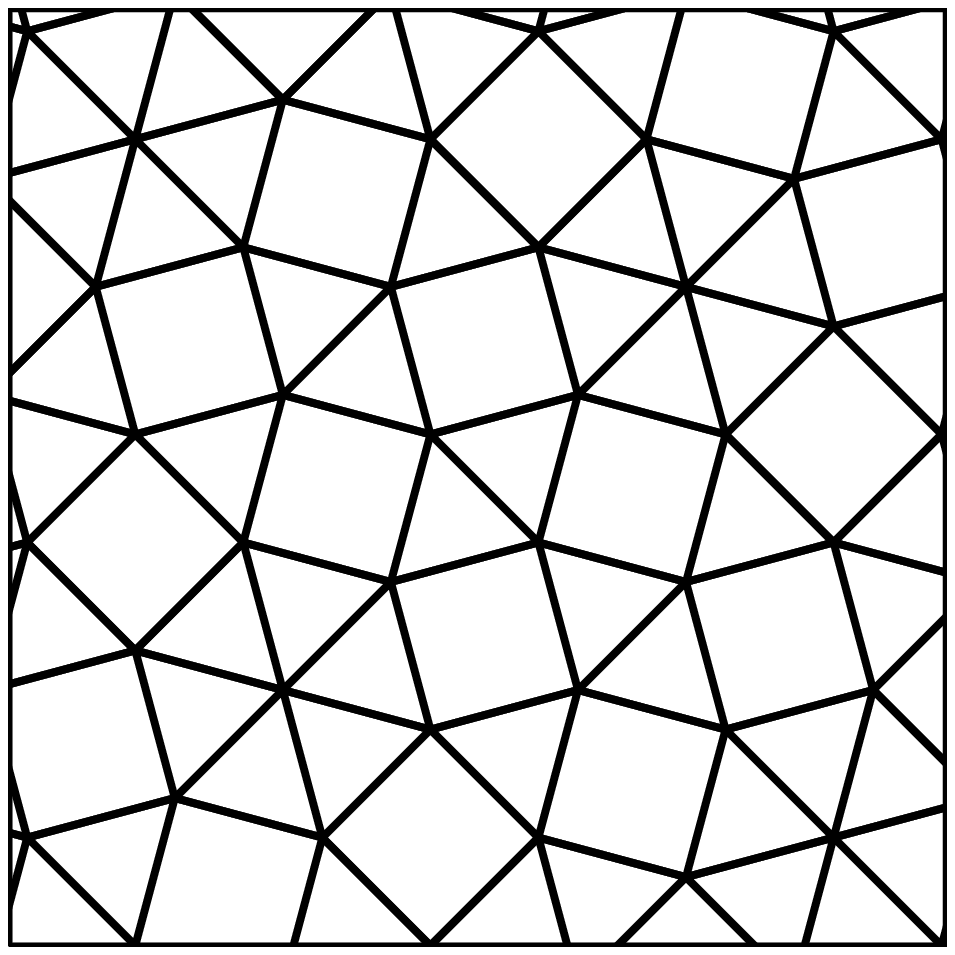} } }
\end{picture} }
\caption{Some examples of configurations of the three solved
rectangle-triangle tilings, the top angle of the triangle is 
$\pi/2$, $2\pi/5$ and $\pi/3$ respectively. \label{rectri}}
\end{figure}

On first sight the lack of an underlying lattice is a great 
difficulty in the analysis.
However this difficulty is not essential as the configurations of the 
tiling can be mapped on a lattice by means of a geometric 
deformation\cite{lattice}.
With this approach Widom studied the tilings of the plane with 
squares and equilateral triangles\cite{widom}. 
Due to the fact that all angles of these objects are 
multiples of $\pi/6$ radians, this model exhibits twelvefold rotational 
symmetry. 
Widom showed that the transfer matrix of the tiling can be diagonalized 
by means of a Bethe Ansatz (BA). The corresponding Bethe-Ansatz 
equations were subsequently solved by Kalugin\cite{kalugin} 
in the thermodynamic 
limit. This solution led to an exact expression of the extensive part 
of the entropy of the tiling.
Later De Gier and the present author have found two other cases
\cite{oct,dec} of 
quasicrystalline random tilings which can be solved by similar techniques. 
The tiles are rectangles and isosceles triangles, immediate 
generalizations of the squares and equilateral triangles.
The rotational symmetry of the maximally symmetric phase 
is different, namely octagonal and decagonal respectively.
Some example configuration of these three solved tilings 
is shown in Figure \ref{rectri}. 
The triangular tile has top angle $\alpha = 2\pi/n$,
where the integer $n$ takes the values 4, 5 and 6.
The rectangle is simply defined to have sides matching in length 
with the legs (length 1) and the base (length $2 \sin \alpha/2$) 
of the triangle.

An obvious question at this point is in what way these tilings are
different from other rectangle-triangle tilings, 
in which the angle $\alpha$ is another rational or even irrational
fraction of $2\pi$.
To see this it is necessary to consider in more detail the geometry 
of the tiling. 
Consider a vertex of the tiling, where $i$ triangles meet with their top 
angle $\alpha$, $j$ triangles with their base angle, and $k$ 
rectangles. Then obviously 
\be i\alpha+j{\pi-\alpha\over 2}+k{\pi\over 2} = 2\pi, \label{vertex} \ee 
and the sum $j+k$ is even. 
Irrespective of the value of $\alpha$ three solutions 
of these conditions 
always exist, namely $(i,j,k) = (0,0,4)$, $(1,2,2)$ and $(2,4,0)$.
Tilings with only vertices of these types we will call generic 
tilings.
In all configurations of these tilings it is possible to vary 
continuously the angle $\alpha$, while the base angle and the 
length of the edges vary accordingly. Therefore the angle $\alpha$ plays no
role in counting the number of ways the plane can be tiled.
However, the restriction to this limited set of vertex types is so 
severe that these tilings have zero entropy: any finite region can be 
tiled in at most one way. For this reason these models are 
not of great interest from the view point of statistical physics.

The tilings are possibly more interesting when $\alpha$ is chosen such 
that other combinations for $(i,j,k)$ are possible.
Though there are many ways to allow for other solutions of 
(\ref{vertex}), there are other conditions to cope with.
The total number of base angles of an entire configuration is always 
twice the number of top angles. Therefore when there is a vertex in 
which $2i<j$ there must in the same configuration also be a vertex with 
$2i>j$.
Therefore the only values of $\alpha$ giving rise to other than generic 
tiling configurations, are those which admit solutions of (\ref{vertex}) 
both with $2i<j$ and $2i>j$. The task to find these values of $\alpha$ is
elementary, but tedious. The result is that $\alpha = 2\pi/n$ with
$n=3$, 4, 5 or 6. All other values of $\alpha$ give only generic 
tilings. The cases $n=4$, 5 and 6 are mentioned above and are 
precisely those tilings that have been solved by means of the Bethe 
Ansatz\cite{kalugin,oct,dec}. 
The case $n=3$ has not been discussed in the literature before. 
The tiles, a triangle with sides 1, 1 and $\sqrt{3}$, and a 
rectangle with sides 1 and $\sqrt{3}$ are such that the vertices of 
this tiling together with the mid-points of the rectangles 
form precisely the sites of a triangular 
lattice. It is tempting to believe that this tiling is also solvable by 
BA, but this does not appear to be the case. 

It may still be true that the rectangle-triangle tilings with
$n=4$, 5 and 6 are 
members of an infinite sequence of solvable models. 
It may be that 
for higher values of $n$ 
the two tiles, the triangle and the rectangle do not suffice, and one 
may need to introduce more tiles as $n$ increases. On the same token 
it may be that for $n=3$ the two tiles are already 
too many, and one should work only with the triangle. 
This indeed gives a solvable tiling with finite entropy: when there
are no rectangles pairs of triangles always share their long side,
and thus form a rhombus. This rhombus tiling has been studied in many
guises\cite{afis}.

\section{Integrability of the square-triangle tiling}
In almost all cases where the BA approach to the 
diagonalization of a transfer matrix or quantum Hamiltonian is 
effective, these operators are members of a commuting family\cite{bb}
The commutativity of this family is proven by the fact that the local 
Boltzmann weights satisfy the Yang-Baxter (YB) equation\cite{bb}.
Such a connection to a YB structure is not apparant for 
the solvable random tilings.
However, in the case of the square-triangle tiling, a connection has 
been found\cite{intst}. 
It turns out that the square-triangle model is equivalent 
to a limit of a known vertex model (associated with the affine Lie 
algebra A$^{(1)}_2$), provided with fields of the 
Perk-Schultz\cite{PS} type. This vertex model does solve the YB equation.
Without repeating the complete argument I will summarize the connection,
and make some comments for later reference. 
For the derivation the reader is referred to\cite{intst}.

The Boltzmann weights denoted as 
$W(\alpha,\beta;\gamma,\delta)$,
with as successive arguments the states of the 
(left, bottom; top, right) legs of the vertex. 
\bea
W(\mu,\mu;\mu,\mu) & = & X_{\mu}^2 \sinh(u+\lambda)\nonumber \\
W(\mu,\nu;\mu,\nu) & = & X_{\mu} X_{\nu}{\rm e}^{u \sgn(\nu-\mu)} \sinh{\lambda} \\
W(\mu,\nu;\nu,\mu) & = & X_{\mu}^2 (Y_{6-\mu-\nu})^{2\sgn(\mu-\nu)}\sinh{u}
\nonumber \eea
where $\mu$ and $\nu \neq \mu$ take the values 1, 2 and 3.
The limit involves both the spectral parameter $u$ and the 
field parameters $X_{\mu}$ and $Y_{\mu}$. 
The square-triangle model is recovered when the 
limit is taken in two steps. In the first step the spectral parameter is 
taken to $u=-\lambda$, at which some of the vertex weights
vanish. 
This is a point of special symmetry where 
the vertices of the square lattice can be factorized 
in vertices of the hexagonal lattice\cite{resh}.
In fact both the spatial symmetry of the hexagonal lattice and the
internal permutation symmetry of the three vertex states are fully observed.
At this point the model is equivalent to the three-coloring 
problem of the edges of the honeycomb lattice such that equally colored 
edges never meet in a vertex. This problem was solved by 
Baxter\cite{3color}. 
In the second step a combination of the Perk-Schultz fields is 
taken to an extreme limit. 
$ X_1^{-1}=X_2=X_3=Y_1^{-1}=Y_2^{-1}=Y_3^{-1}=x$ and take $x \to 0$. 
In this limit a few of the remaining Boltzmann 
weights vanish as well. 
Because in the first limit the spectral parameter as a
free variable is lost, so is also the YB structure.
In order to retain the spectral parameter one might wish to take the 
field limit first.
However in that procedure one of the vertex weights would diverge,
or after suitable normalization, be the only one to survive.
Thus we see the two limits do not commute.
As a consequence we come to the counter-intuitive conclusion
that the Boltzmann weights of the 
square-triangle tiling are a limit of a solution to the YB equation,
but these weights themselves are not a proper solution.
It thus appears that the solvability of square-triangle model is related 
to the YB equation, albeit somewhat remotely.

In the remainder of this paper we investigate in more detail the solvable 
rectangle-triangle tiling with octagonal symmetry, and in particular discuss 
the question if it also can be viewed as a limit of a YB-solvable model.
For this purpose we review the solution, 
though in an approach
different from that published before\cite{oct}. 
Here we try to keep the symmetry of the tiling as much as possible.

\section{The octagonal rectangle-triangle tiling}
\begin{figure}
\centerline{\epsfxsize=11cm \epsfbox{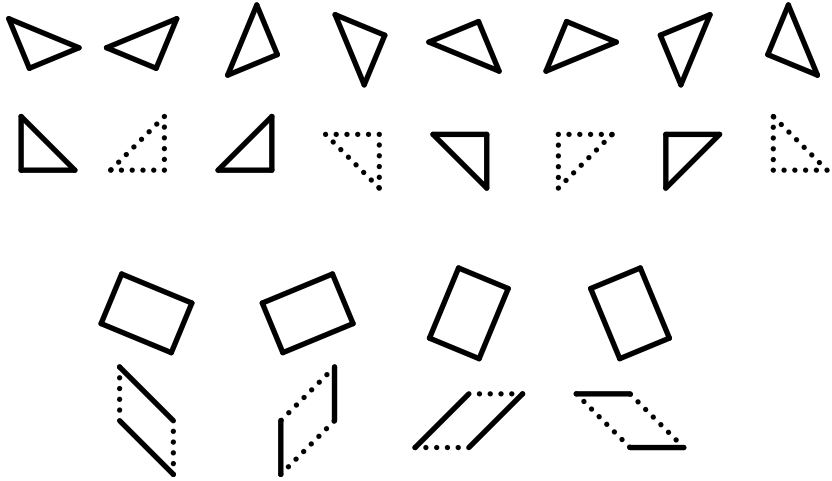}}
\caption{Deformations of the triangular and rectangular tiles in their 
respective orientions. \label{tiles}}
\end{figure}

This model 
was first considered by Cockayne\cite{cock} in a slightly different language.
The plane (or a finite periodic section of it) 
is tiled with rectangular, isosceles triangles 
and rectangles of which the sides, length 1 and $\sqrt{2}$, 
match those of the triangle. One may, as in a canonical 
ensemble fix the relative density of each of the tiles, or as in the grand 
canonical ensemble control the relative density by means of an activity 
variable.
The tiles are not permitted to overlap, and may leave no space uncovered.
Since all the angles in the tiles are multiples of $\pi/4$, the angle 
between any two edges in a tiling configuration must also be such 
multiple. Therefore the rectangles may occur in four, and the triangles 
in eight different orientations, shown in figure \ref{tiles}. 

As a first step in the analysis the tiles are deformed, such that they 
are commensurable with a regular square lattice but 
continue to cover the plane without holes or overlap.
The deformation is illustrated in figure \ref{tiles} and may be described 
as follows. Let all the edges in the original tiling make an angle with 
the horizontal equal to an odd multiple of $\pi/8$ radians.
The short edges are rotated over $\pi/8$ to the left or to the right, 
such that they end up horizontally or vertically. The long edges are also 
rotated over $\pm \pi/8$ but such that they end up in one of the diagonal 
directions. As a result the triangles are rotated rigidly and the 
rectangles are deformed into parallelograms.
The deformed tiles fit precisely on the square lattice, and any tiling
configuration can be viewed as the state of a lattice model.
The tile edges in the resulting lattice are marked by either solid or 
dotted lines to distinguish the original orientations. 
This is necessary because two differently oriented edges are mapped onto 
the same orientation.
The fact that the mapping, while it deforms the entire configuration,
does not create any holes or overlaps, follows directly from the 
observation that it is defined as a mapping of the edges, i.e. the 
shared boundary between adjacent tiles.

The configurations of the original tiling are mapped one-to-one on 
configuratings of the lattice tiling.
Therefore the combinatorial problem of 
counting the number of possible tilings seems unaltered.
However, because the overall shape and the area of a tiling is altered 
by the mapping, the problem of tiling a given section of the plane with 
a given number of tiles is not the same. 
In the thermodynamic limit this distinction will be insignificant.
The change of area can be accounted for, and the shape does not
matter for the thermodynamic functions.

\begin{figure}
\centerline{\epsfxsize=11cm \epsfbox{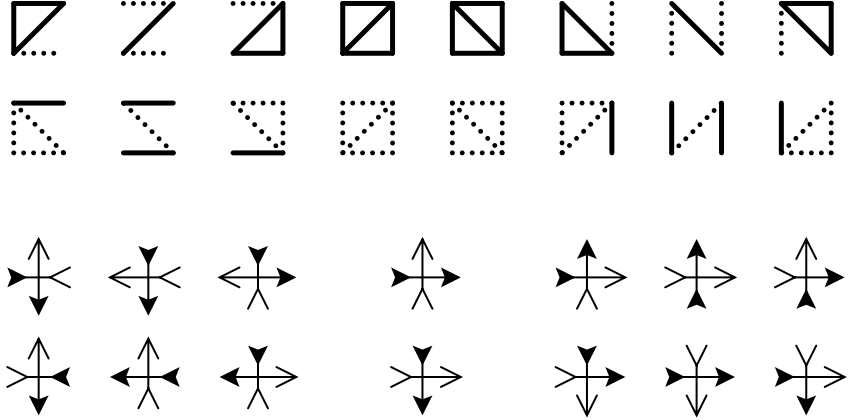}}
\caption{The top two lines show the possible configurations of an
elementary square of the lattice model corresponding to the 
octagonal tiling model. The bottom two  lines show the corresponding
configurations of the fourteen-vertex model. \label{plaq}}
\end{figure} 

The lattice representation of the tiling can be compactly described by 
listing the possible states of each elementary square of the lattice, as 
shown in figure \ref{plaq}. 
The sides of the elementary squares can be in one of four states, and
represent either a short edge of the tiling, or a diagonal of a 
rectangle. The local configurations of the elementary faces of the 
lattice can be represented as fourteen different states 
of a vertex model which has four possible states for each edge.
Figure 
\ref{plaq} shows a representation in terms of two types of arrows,
which will be called open and filled. 
This notation makes immediately evident two conservation laws, 
the net flux of open and that of filled arrows.
We will refer to this lattice model as fourteen-vertex model.

\section{Bethe Ansatz equations}
A coordinate BA formalism can be set to diagonalize 
the vertex model shown in figure \ref{plaq}. 
For the reference state we choose the open
arrow up. The quasi-excitations are the locations with a vertical 
filled arrow, up or down. 
The open arrow down serves as a bound state between 
these two different excitations.
Besides the conservation of open and filled flux, there is an
additional conservation law, which is less evident. It turns out that
at every application of the transfer matrix the excitations move one
step either to the left or to the right. Therefore the lattice can be
divided into two sublattices such that the number of excitations on
each sublattice remains constant from row to row.
As a result there are in total three conserved number and 
correspondingly three families of BA variables.

In order to find the most general solution the vertex weights of the 
model are kept general.
As the BA eigenstate takes form, however, several restrictions on
the weights are necessary.
In the first place it turns out that the weights of the vertices 
almost factorize into weights associated with the tiles.
Complete factorization would imply completely non-interacting tiles
(apart from the ban on holes and overlaps).
The one exception is that where two triangles form a square by 
sharing their long edge, this configurations has an extra weight
$1/2$.
The tiles may have different weights in the different orientation, but
there are restrictions on these weights.
We denote the weights of the oriented triangles as $t_j$, with
$j=1,\cdots,8$ and of the rectangles as $r_j$, with $j=1,\cdots,4$,
both in the order in which they are shown in figure \ref{tiles}.
Then the restrictions are:
\begin{eqnarray} 
t_1 t_5 & = & t_3 t_7 \nonumber \\
t_2 t_6 & = & t_4 t_8 \label{tiles.restr}\\
r_1 r_3 & = & \pm r_2 r_4  \nonumber
\end{eqnarray}
Because we favor a statistical interpretation of the tiling 
we choose the positive sign in the last line. 

The resulting BA equations are, for a lattice with even horizontal 
size $L$, and periodic boundary conditions.
\bea
x_p^L  & = & (-)^{n_x+1} A \prod_{k=1}^{n_{\lambda}}  \left(\lambda_k -  x_p^2\right)
\nonumber \\		 		        
y_q^L  & = & (-)^{n_y+1} A \prod_{k=1}^{n_{\lambda}}  \left(\lambda_k -  y_q^2\right) 
\label{tilba} \\
1 & = & (-)^{n_{\lambda}+1} B  \prod_{p=1}^{n_x}  \left(\lambda_k -  x_p^2\right) 
 \prod_{q=1}^{n_y}  \left(\lambda_k -  y_q^2\right) 
\nonumber
\eea 
In terms of the solutions $x_p$, $y_q$ and $\lambda_k$ of these
equations the eigenvalue of the transfer matrix reads
\be \Lambda = C \prod_{k=1}^{n_{\lambda}} \lambda_k
\; \prod_{p=1}^{n_x} x_p^{-1}
\; \prod_{q=1}^{n_y} y_q^{-1}
\ee
The constants are expressed in the tile weights:
\bea
A & = & \left({t_1 t_5 r_3 \over t_2 t_6 r_4}\right)^{n_{\lambda}}
\nonumber \\
B & = & \left({t_1 t_5 \over r_4}\right)^{L}
 \left({t_1 t_5 r_3 \over t_2 t_6 r_4}\right)^{n_x + n_y}
\\
C & = & (t_1 t_5)^{L} 
\left({r_1\over t_1 t_5}\right)^{n_x + n_y - n_{\lambda}}
\left({r_2 \over r_1}\right)^{n_{\lambda}}
. \nonumber \eea
The form of these BA equations is typical for tiling
models. 
The eigenvalue expression is a simple product rather than a
sum of, say four, products, according to the number of edge states.
The factors in the BA equations are 
binomial rather than rational or trigonometric.

\section{Search for a Yang-Baxter structure}
Even though the tiling model has a number of free parameters 
in the solution,
none of those plays the role of a
spectral parameter. The only way that the parameters feature in the
eigenvalue is via the combination $C$ as an overall factor.
As the BA equations are the most general given the 
set of vertex configurations with non-zero weight, 
the only way to introduce a spectral parameter is to include other 
vertex configurations.
The natural choice is to include only configurations that satisfy the
flux conservation of both types of arrows.
Our attempts in this direction have all failed, as they led only to
null solutions.
Incapable of an exhaustive search, we can not make strong statements
about this possibility.

However, the situation may be similar to the case of square-triangle
model, described above.
If that is so it would be futile to seek solutions of the YB
equations which include the fourteen-vertex model. Instead this model
would be a singular limit of such a solution, and not a true member.
It is on this assumption that we proceed. 
Guided by the results concerning the integrability of the 
square-triangle model, we 
seek a vertex model of which the vertex configurations include those
of the fourteen-vertex model, but have a higher symmetry. 
This would still have a fixed value of the spectral parameter, 
and permit the introduction of field parameters, 
such that in a special limit the fourteen-vertex model is recovered.

A natural choice of a symmetrized version of the fourteen-vertex model
is extending the permitted vertex states with those obtained by 
arrow inversion from the original fourteen. 
The resulting set of vertices are twenty-four states in which one open
and one filled arrow enter and leave the vertex.
We propose that it is possible to assign weights to these twenty-four
vertex states such that 
(i) they form a member of a solution of the YB equation, and 
(ii) permit a limit in which they reduce to the fourteen vertices of the 
octagonal tiling problem.
To be more precise we propose that there exists a solution of the 
YB equation
\be
{\sum_{\mu'',\nu'',\gamma}
W''(\mu,\nu;\nu'',\mu'') 
W(\mu'',\beta;\gamma,\mu')  
W'(\nu'',\gamma;\alpha,\nu') = } \nonumber \ee \be
{ \sum_{\mu'',\nu'',\gamma}
W''(\mu'',\nu'';\nu',\mu') 
W'(\nu,\beta;\gamma,\nu'')
W(\mu,\gamma;\alpha,\mu''),  } \label{yb}
\ee
such that the Greek symbols take the four arrow states as values.
Part (i) of the proposal states that
the symbol $W$ is non-zero only for the twenty-four cases of its 
arguments when one open and one filled arrow enter and leave the 
corresponding vertex. 
Guided by the YB structure of the square-triangle model, we expect that the 
symbols $W'$ and $W''$ may have
nonzero entries besides these ones.
Part (ii) of the proposal states that the solution manifold permits a 
limit to be taken in which the symbol $W$ is reduced to weights of 
the fourteen 
vertices corresponding to the octagonal tiling model (figure \ref{plaq}).
In this limit some elements of the symbols $W'$ or $W''$ may well diverge.

Before discussing tests of this proposal 
we make a few remarks concerning the symmetry of the model.
Consider a configuration of these vertices on a square lattice.
The edge states are denoted by the kind of arrow, open or filled, and by 
the direction, up or down for vertical edges and left or right for 
horizontal ones.
If the sites of the lattice are divided into two square sublattices A and B, 
the direction of the arrows can be denoted as A to B or vice versa.
With this labelling, the four edges incident in the same vertex are 
always in different states. In other words the configurations of the 
24-vertex model are in one-to-one correspondence to the colorings of 
the edges 
of the square lattice with four colors, such that in no vertex 
two edges of the same color meet. Clearly in this coloring problem the 
four colors can be freely permuted without altering the ensemble of 
coloring configurations.
Since any of the twenty-four vertex configurations can be turned into 
any other one by a suitable permutation of the four edge states, the 
only weight assignment invariant for these permutations gives all 
vertices equal weight.
It is likely that if there is an integrable manifold in this 24-vertex 
model, it includes the symmetric point where all weights are equal,
equivalent to the four-coloring problem of the edges of the square 
lattice.

In order to test the above proposal we set up and attempt to solve the 
YB equations (\ref{yb}). 
The most restrictive and simple approach is to insert for all three 
symbols, $W$, $W'$ and $W''$ just the twenty-four weights described above, 
setting all other weights equal to zero. 
This turns out to lead only to a trivial null solution.
Thus, the full solution, if it exists reuires more than the twenty-four 
vertices. A natural extension is with those vertices in which the total 
flux of each types of arrows is still conserved.
Those vertex configuration are the well-known six-vertex configuration, 
now in two types, those with only open arrows and those with only filled 
arrows, twelve configurations in total.
Together with the original twenty-four vertices this makes thirty-six.
We have not succeeded in solving the YB equations for these thirty-six 
weights for each of the symbols $W$, $W'$ and $W''$, due to the 
complexity of the problem.
However, by allowing for $W$ only the original twenty-four weights
while retaining the full thirty-six weights for $W'$ and $W''$, the 
the problem is considerably simpler and could in fact be solved.
Unfortunately, the resulting weights for $W$ can not be made all 
positive. Therefore, 
contrary to our expectation the four-coloring problem is not included in 
the solution. 
Even so, part (i) of the proposal is verified, albeit with some negative
weights. We note that
the twenty-four weights can be made equal in absolute value, even though 
they can not all be made positive.

The solution of this reduced set of YB equation has still a great number 
of free parameters. 
This freedom permits the reduction of the weights $W$ to only the 
fourteen non-zero weights of the octagonal tiling model. 
Interestingly, in that limit the same restriction on the  
vertex weights is found as those following from the BA,
with the sign in the third line of 
(\ref{tiles.restr}) no longer free, but negative.
In other words, one, or three of the weights of the oriented rectangles 
must be negative. 
Thus we have not found the YB structure behind the integrability of the 
octagonal tiling problem, but something intriguingly close to it.
We can not recover the tiling problem from a YB structure, but then the 
tile weights can not be all positive.
Thus part (ii) of the proposal is almost verified: the
solution to the YB equation has a limit which up to signs reproduces the 
weights of the tiling model.

\section{The four-coloring problem}
In the previous section we encountered the combinatorial problem of 
coloring the edges of the square lattice in four colors, such that 
two equally colored edges never meet in the same vertex.
This four-coloring problem has been 
studied also in its own right\cite{4color}.
It is related to fully packed loop models on the square 
lattice\cite{fpl} which has application to physics of polymers in 
the melt\cite{melt}. It is also related to the Hamiltonian walk problem
\cite{hamwalk}.

As described above, we have not succeeded in finding a solution to the 
YB equation which includes this model, at least not the coloring problem 
in which all configurations have positive weight.
This does not imply that the model is not integrable.
To yet have an indication of its integrability we have 
attempted to construct BA eigenectors to the transfer matrix.
We used the formulation of the 24-vertex model. 
Thus the edges of the square lattice wrapped on a cylinder of 
even circumference $L$ all carry an open or filled arrow, 
pointing along the edge. The configurations in which one 
arrow of each type points into each vertex have weight one, all other 
configurations have weight zero. 

The coordinate (nested) BA approach indeed yields eigenvectors 
to the transfer matrix, rather similar in structure 
to those of the octagonal tiling model (\ref{tilba}). 
With respect to a reference state, all open arrow up, the excitations 
are the filled arrows. The open arrow down serves again as a 
bound state of two opposite filled arrows.
With every application of the transfer matrix the excitations
that do not form a bound state, move a one step, to the right or to the 
left. As a consequence not only the number of excitations is conserved, 
but also their distribution over two sublattices.

Precisely as in the fourteen-vertex model,
two families of variables give the momenta of the excitations on the
even and odd sublattice respectively, and a third set of variables is 
associated with the distribution of the filled down arrows among all of 
the filled arrows.

The resulting BA equations read
in suitable variables
\[
\left({ 1 + u_j \over 1 - u_j}\right)^{L/2}
 =  -\left(\prod_{m=1}^{n_w} { w_m - u_j + 1 \over u_j - w_m + 1}\right)
   \left(\prod_{k=1}^{n_u} { u_j - u_k + 2 \over u_j - u_k - 2}\right) 
\] \be
\left({ 1 + v_j \over 1 - v_j}\right)^{L/2}
 =  -\left(\prod_{m=1}^{n_w} { w_m - v_j + 1 \over v_j - w_m + 1}\right)
   \left(\prod_{k=1}^{n_v} { v_j - v_k + 2 \over v_j - v_k - 2}\right)
\label{4cba} \ee \[
-1  = 
\left(\prod_{j=1}^{n_u} { w_m - u_j + 1 \over u_j - w_m + 1}\right)
\left(\prod_{k=1}^{n_v} { w_m - v_k + 1 \over v_k - w_m + 1}\right)
\left(\prod_{l=1}^{n_w} {w_l - w_m + 2 \over w_l - w_m - 2}\right)  
\]
The eigenvalue can be written in terms of the solutions of these 
equations
\bea \Lambda & = &
\left(\prod_{j=1}^{n_u} { 1 + u_j \over 1 - u_j}\right)^{1/2} 
\left(\prod_{k=1}^{n_v} { 1 + v_k \over 1 - v_k}\right)^{1/2} 
\left(\prod_{m=1}^{n_w} {2 - w_m \over w_m}\right) \nonumber \\ & + &
\left(\prod_{j=1}^{n_u} { 1 - u_j \over 1 + u_j}\right)^{1/2} 
\left(\prod_{k=1}^{n_v} { 1 - v_k \over 1 + v_k}\right)^{1/2} 
\left(\prod_{m=1}^{n_w} {- 2 - w_m \over w_m}\right) \label{4cev}
\eea
These equations have been derived for the sectors 
where the excitations are relatively sparse.
We have no proofs for these equations in full generality.
We have numerically verified them for arbitrary sectors 
of lattices up to twelve sites in the horizontal direction.
When the roots of the equation include a $w_m=0$ the eigenvalue is 
undetermined from these expressions.
This ambiguity could be resolved
if we could find the eigenvalue expression with a spectral parameter.
Altogether the BA approach shows that indeed this
four-coloring problem is integrable. 

These BA equations look very much like those derived by 
Martins\cite{mixed} for mixed SU($N$) vertex models for the case $N=4$.
However, the sign of some of the factors is different
(those involving both $u$ and $w$ or both $v$ and $w$). 
The eigenvalue (\ref{4cev}), 
aside from the absence of a spectral parameter here, 
differs from also from that of Martins 
in an overall factor, which depends on 
the BA roots. 
In fact the comparison inspires to introduce a spectral parameter $u$ in our 
expression for the eigenvalue. 
This process may seem arbitrary, but it is
highly constrained for the following consideration.
It is known\cite{bb} that the BA equations follow directly from the 
expression for the eigenvalue by the requirement that
the eigenvalue be an entire function of the spectral parameter.
Here we follow the reverse argument, we know the BA equations and 
the eigenvalue for one value of the spectral parameter,
and propose a specific dependence on the spectral parameter
such that the poles from the different terms in the expression 
cancel against each other as a consequence of the BA 
equations. 
\[ \Lambda(u)  = 
\left(\prod_{j=1}^{n_u} { 1 + u_j\over 1 - u_j}\right)^{1/2} 	
\left(\prod_{j=1}^{n_v} { 1 - v_j\over 1 + v_j}\right)^{1/2} \times
\] \[   \left[  (1-u^2)^{L/2}
\left(\prod_{j=1}^{n_v} { 1 + v_j-2u \over 1 - v_j+2u}\right)
\left(\prod_{j=1}^{n_w} {2 - w_j+2u \over w_j-2u}\right) \right.
\] \[
 +  (1-u^2)^{L/2}
\left(\prod_{j=1}^{n_u} { 1 - u_j +2u\over 1 + u_j-2u}\right)
\left(\prod_{j=1}^{n_w} {- 2 - w_j+2u \over w_j-2u}\right) 
\] \[
 + 
(u+u^2)^{L/2}
\left(\prod_{j=1}^{n_u} \frac{-3-u_j+2u}{1+u_j-2u}\right)
\] \be  \left. + (u^2-u)^{L/2}
\left(\prod_{j=1}^{n_v} \frac{-3+v_j-2u}{1-v_j+2u}\right) \right]
\label{analytic.ba} \ee
The value $u=0$ corresponds to the original
expression (\ref{4cev}).
One check of this proposal is that it now allows us to 
calculate the eigenvalue unambiguously also in those cases where one of 
the $w_m$ roots vanishes. 
Indeed it turns out that these cases the limit $u \to 0$ of the 
expression correspond to eigenvalues of the
transfer matrix that were undetermied before.

The eigenvalue expression (\ref{analytic.ba}) together with the BA 
eigenvectors of the transfer matrix of the 24-vertex model
unambiguously defines a matrix.
If this matrix can be built up from local Boltzmann
weights is yet to be verified. 
The whole expression has some unusual features, such as the obvious
overall factor, independent of the spectral parameter, and involving a 
square root.

\section{Summary}
Quasicrystals show rotational symmetries which are forbidden by the 
rules of crystallography. This is possible because quasicrystals are 
aperiodic. 
From the fact that quasicrystals exist we must conclude their structure 
is a minimum of the free energy. 
It can be argued that in quasicrystals, more than in crystals, the 
entropy is, compared with the energy, 
plays a significant role in determining this minimum.
For these materials random tilings serve as model systems. 

This paper reviews some results of two-dimensional random tilings 
of which the entropy has been calculated exactly.
These models have phases with 
octagonal, decagonal and dodecagonal spatial rotational symmetry. 
They appear as three member of an infinite sequence, but there
properties, among which solvability, 
uniquely set them apart from the rest of the sequence.
We have discussed the question why this is the case.

While these models have been solved by means of the Bethe Ansatz, 
an underlying Yang-Baxter structure is not apparent.
Only in the dodecagonal case the relation with a solution to the YB
equation has been known for a few years. 
Here the connection involves a model family which includes 
the coloring problem of the edges of the hexagonal lattice
with three colors, such that the edges meeting in a vertex have
different colors. The tiling model is a singular limit of this family.

This paper takes a few steps at finding 
a similar connection for the solved tiling
model with octagonal symmetry. 
Its transfer matrix also permits a BA solution. 
The search for the YB structure behind this integrability is not 
completed.
However, the trace seems to lead, just as in the dodecagonal case, by 
a coloring problem. In this case it is the problem of coloring the edges 
of the square lattice in four colors such that nowhere two edges with 
the same color meet in a vertex. 
We give the BA equations of this problem, thus showing that the model is 
integrable.
\section*{Acknowledgments}
The author wishes to thank the organizers of
``Baxter's revolution in mathematical physics'', and
R. Tateo, S.O. Warnaar, J.C. de Gier and 
D. dei Cont for many valuable discussions.



\begin{thebibliography}{99}
\bibitem{penrose}
R. Penrose,
 The role of aesthetics in pure and applied mathematical research,
 {\em Bull. Inst. Math. Appl.} {\bf 10} (1974), 266.
\bibitem{ammann}
L.J. Shaw, V. Elser, and C.L. Henley, 
Long-range order in a three-dimensional random-tiling quasicrystal
{\em Phys. Rev.}~B {\bf 43} (1991), 3423

\bibitem{elser}
V.~Elser, 
Comment on ``Quasicrystals: a new class of ordered structures".
{\em Phys.~Rev.~Lett.}  {\bf 54} (1985), 1730, and
Indexing problems in quasicrystals diffraction. 
{\em Phys.~Rev.} B {\bf 32} 1985), 4892.
\bibitem{rt1}
L.H. Tang,
Random-tiling quasi-crystal in 3 dimensions, 
{\em Phys. Rev. Lett.} {\bf 64} (1990), 2390
\bibitem{rt2}
K.J. Strandburg,
Random-tiling quasicrystal,
{\em Phys. Rev.} B {\bf 40} (1989), 6071
\bibitem{rtl}
Richard C, Hoffe M, Hermisson J, M. Baake,
Random tilings: concepts and examples, 
     {\em J. Phys.} A {\bf 31} (1998), 6385
\bibitem{dimers}
P.~W. Kasteleyn,
The statistics of dimers on a lattice,
\newblock {\em Physica} {\bf 27} (1961), 1209,
\newblock and H.~N.~V. Temperley and M.~E. Fisher,
\newblock {\em Phil. Mag.} [8th Ser.] {\bf 6} (1961), 1061.
\bibitem{hh}
R.J. Baxter, Hard hexagons: exact solution, {\em J. Phys.}~A
{\bf 13} (1980), L61
\bibitem{lattice}
 C.L.~Henley, 
Random tilings with quasicrystal order: transfer-matrix 
approach, {\em J.~Phys.}~A {\bf 21} (1988), 1649.
\bibitem{widom}
M.~Widom,
\newblock {B}ethe {A}nsatz solution of the square-triangle random tiling model.
\newblock {\em Phys. Rev. Lett.} {\bf 70} (1993), 2094.
\bibitem{kalugin}
P.A. Kalugin.
\newblock The square-triangle random-tiling model in the thermodynamic limit.
\newblock {\em J. Phys.}~A {\bf 27} (1994), 3599.
\bibitem{oct}
J.~de~Gier and B.~Nienhuis, Exact solution of an octagonal tiling 
model, {\em Phys.~Rev.~Lett.} {\bf 76} (1996), 2918-2921 and
J.~de~Gier and B.~Nienhuis.
\newblock The exact solution of an octagonal rectangle-triangle random tiling.
\newblock {\em J. Stat. Phys.} {\bf 87}, 415.
\bibitem{dec}
J.~de~Gier and B.~Nienhuis.
\newblock Bethe Ansatz solution of a decagonal rectangle-triangle random 
tiling,
\newblock {\em J.~Phys.}~A {\bf 31} (1998), 2141-2154.
\bibitem{afis}
H.W.J.~Bl\"ote and H.J.~Hilhorst, Roughening transitions and the
zero-temperature triangular Ising antiferromagnet,
{\em J.~Phys.}~A {\bf 15} (1982), L631
\bibitem{bb}
R.J.~Baxter, {\it  Exactly Solved Models in Statistical Mechanics},
Academic Press, London (1982).
\bibitem{intst}
J.~de~Gier and B.~Nienhuis,
\newblock Integrability of the square-triangle random tiling model.
\newblock {\em Phys. Rev.} E {\bf 55} (1997), 3926.
\bibitem{PS}
J.H.H. Perk and C.L. Schultz, New families of commuting transfer
matrices in $q$-state vertex models, {\em Phys. Lett.} {\bf 84A} (1981), 407
\bibitem{resh}
N.Y. Reshetikhin, 
A new exactly solvable case of an O($n$) model on a hexagonal lattice
{\em J. Phys.} A {\bf 24} (1991), 2387
\bibitem{3color}
R.J. Baxter, Coloring of a hexagonal lattice, {\em J. Math. Phys.} 
{\bf 11} (1970), 784 
\bibitem{cock}
E. Cockayne, Atomistic octagonal random-tiling model, {\em J. Phys.}
A: Math. Gen. {\bf 27} (1994), 6107
\bibitem{4color}
J. Kondev and C.L. Henley, 
Four-coloring model on the square lattice - a critical ground-state,
{\em Phys. Rev.} B {\bf 52} (1995), 6628
\bibitem{fpl}
J.L. Jacobsen,
\newblock On the universality of fully packed loop models,
{\em J. Phys.} A {\bf 32} (1999), 5445
and   S. Higuchi,
\newblock Compact polymers on decorated square lattices
{\em J.Phys.} A {\bf 32} (1999), 3697
\bibitem{hamwalk}
M.T. Batchelor, H.W.J. Bl\"ote, B. Nienhuis and C.M. Yung,
Critical behavior of the Fully Packed Loop model on the square 
lattice, {\it J. Phys.} A {\bf 29} (1996), L399
\bibitem{melt}
J.L. Jacobsen and J. Kondev,
Field theory of compact polymers on the square lattice,
{\em Nucl. Phys.} B {\bf 532} (1998), 635, 
and
J. Kondev  and J.L. Jacobsen,
Conformational entropy of compact polymers, 
{\em Phys. Rev. Lett.} {\bf 81} (1998), 2922
\bibitem{mixed}
M.J. Martins, 
\newblock Integrable mixed vertex models from braid monoid algebra,
\newblock (solv-int/9903006) in {\em Statistical physics on the eve of
the 21-st century}, eds. M.T. Batchelor, L.T. Wille, Vol 14 of 
{\em Series on anvances in Statistcal Mechanics}, World Scientific,
Singapore 1999.
\end{thebibliography}
\end{document}